\documentstyle[PASJadd, psfig]{PASJ95}
\newcommand{\vol}{999}
\newcommand{\no}{1}
\newcommand{\lett}{}
\newcommand{\spage}{1}
\newcommand{\rdate}{1999 September 30}
\newcommand{\adate}{1999 November 30 }

\begin{document}

\title{High-Resolution Near-Infrared Imaging of the Powerful Radio Galaxy 3C 324 at $z$ = 1.21 with the Subaru Telescope}

\author{Toru {\sc Yamada}, Masaru {\sc Kajisawa}, Ichi {\sc Tanaka} \\
{\it Astronomical Institute, Tohoku University, Aoba-ku, } \\
{\it Sendai, Miyagi 980-8578} \\
{\it E--mail(TY): yamada@astr.tohoku.ac.jp } \\
 \\
Toshinori {\sc Maihara}, Fumihide {\sc Iwamuro}, Hiroshi {\sc Terada}, Miwa {\sc Goto}, Kentaro {\sc Motohara}, \\
Hirohisa {\sc Tanabe}, Tomoyuki {\sc Taguchi}, Ryuji {\sc Hata} \\
{\it Department of Physics, Faculty of Science, Kyoto University, Sakyo-ku,} \\
{\it Kyoto 606-8502} \\
 \\
Masanori {\sc Iye}, Masatoshi {\sc Imanishi}, Yoshihiro {\sc Chikada} Michitoshi {\sc Yoshida} \\
{\it National Astronomical Observatory, 2-21-1, Osawa, Mitaka,} \\
{\it  Tokyo 181-8588, Japan} \\
 \\
Chris {\sc Simpson}, Toshiyuki {\sc Sasaki}, George {\sc Kosugi}, Tomonori {\sc Usuda} \\
Koji {\sc Omata}, Katsumi {\sc Imi} \\
{\it Subaru Telescope, National Astronomical Observatory of Japan,} \\
{\it  650 North Aohoku Place, Hilo, HI 96720, U.S.A. }\\
\vspace{2.0cm}
 }

\abst{
 We have obtained high-resolution $K^\prime$-band images of the powerful $z=1.206$ radio galaxy 3C 324 with the Subaru telescope under seeing conditions of 0$^{\prime\prime}$.3--0$^{\prime\prime}$.4. We clearly resolved the galaxy and directly compared it to the optical images obtained with the Hubble Space Telescope. The host galaxy of 3C 324 is revealed to be a moderately luminous elliptical galaxy with a smooth light profile. The  effective radius of the galaxy, as determined by profile fitting, is 1.3$\pm$ 0$^{\prime\prime}$.1  (1.2 kpc), which is significantly smaller than the value of 2$^{\prime\prime}$.2 published in Best et al.\ (1998, MNRAS, 292, 758). The peak of the $K^\prime$-band light coincides with the position of the radio core, which implies that the powerful AGN lies at the nucleus of the host galaxy. The peak also coincides with the gap in the optical knotty structures which may be a dust lane hiding the UV-optical emission of the AGN from our line of sight; it is very likely that we are seeing the obscuring structure almost edge-on. We clearly detected the `aligned component' in the $K^\prime$-band image by subtracting a model elliptical galaxy from the observed image. The red $R_{{\rm F702W}}-K$ color of the outer region of the galaxy avoiding the aligned component indicates that the near infrared light of the host galaxy is dominated by an old stellar population.  }
\kword{galaxies: active --- galaxies: evolution
 --- galaxies: elliptical and lenticular, cD --- galaxies: individual:
(3C 324)}
\setcounter{page}{1}
\maketitle
\thispagestyle{headings}

{  \it To be appear in PASJ vol. 52, 2000}


%
\clearpage

\section{Introduction}

 Powerful radio galaxies provide one of the most exciting opportunities to trace the evolution of galaxies and their activity over the history of the Universe. There is a rapid evolution of the power of active galactic nuclei (AGN), such as quasars and radio galaxies, toward $z\sim3$, and such evolution may reflect some aspects of the phenomena of galaxy formation and evolution.  Powerful radio galaxies have an advantage over quasars in studying the structures in hosts since they are not swamped by the bright nucler light.

 One of the most striking properties of high-redshift powerful radio galaxies (HzPRGs) is the `alignment effect' observed in their optical images (McCarthy et al. 1987; Chambers et al. 1987; Rigler et al. 1992). Rest-frame ultraviolet (UV) radiation tends to have an elongated structure aligned with the radio-jet axis.
 The origin of this aligned UV radiation is still not clear; while the scattered non-thermal radiation contributes to the observed UV flux, at least to some extent, in many radio galaxies, the stellar photospheric emission and/or the nebular continuum emission has also been observed in some sources (e.g., Cimatti et al. 1996). In any case, UV emission is very likely to be the {\it consequence} of nuclear activity, and near-infrared observations may be more essential to reveal the true host-galaxy structure.

 3C 324 at $z=1.206$ has been extensively studied as one of the `proto-typical' objects which show the optical alignment effect. Its total radio power at 5 GHz is large, log $P_{\rm 5 GHz}$= 27.75 ($H_0$ = 50 km s$^{-1}$Mpc$^{-1}$, $q_0$ = 0.5; hereafter we use this set of cosmological parameters unless noted), which implies the existence of a very powerful AGN. Indeed, Cimatti et al. (1996) detected a broad Mg {\sc II} $\lambda$ 2800 \AA\ emission line in the polarized spectrum, and thus established that 3C 324 hosts a quasar obscured by dust. The radio source has two distinct radio lobes with a separation of $\sim 11$$^{\prime\prime}$ (95 kpc at $z=1.206$; Fernini et al. 1993; Best et al. 1998b) at a position angle of 71$^\circ$ (Dunlop, Peacock 1993). Best et al. (1998b) detected a plausible radio core at frequencies of 8.2 and 4.7 GHz. Hubble Space Telescope (HST) WFPC2 observations with the ${\rm F702W}$ and ${\rm F791W}$ filters (Longair et al. 1995; Dickinson et al. 1995; Best et al. 1997, 1998a) revealed in detail the elongated patchy structure of the rest-frame UV emission, which is collinear with the radio axis. The radio core is located in a gap between the two central optical knots (Best et al. 1998b), which suggests that this gap is due to a dust lane which obscures the active nucleus. The existence of a significant amount of dust is supported by the detection of sub-mm radiation which is not associated with the synchrotron radiation (Best et al. 1998c).

 At near-infrared (NIR) wavelengths, previous ground-based observations under $\sim$ 1$^{\prime\prime}$ seeing conditions have detected the red host galaxy (Dunlop, Peacock 1993; Dickinson et al. 1995;  Best et al. 1997, 1998a). Dunlop and Peacock (1993) showed that the host galaxy has a somewhat elongated structure whose position angle (PA) is $\sim 75^\circ$, closely aligned with the radio axis. Best et al. (1998a) argued that the galaxy is as bright as $K=16.99$ mag (9$^{\prime\prime}$ aperture) and  has a light profile well represented by a de Vaucouleurs law with an effective radius of 2$^{\prime\prime}$.2 (19 kpc). Dickinson et al. (1995) pointed out that the $K$-band light peak coinsides with the gap of the optical knots. It has been known that 3C 324 is located in a galaxy cluster (Dickinson et al. 1997; Kajisawa et al. 1999 and references therein) and is the brightest galaxy in the cluster at z = 1.21. 

 We have observed 3C 324 with the Subaru telescope equipped with the Cooled Infrared Spectrograph and Camera for OHS (CISCO: Motohara et al. 1998), which provides a $\sim 2$$^{\prime\prime}$ $\times 2$$^{\prime\prime}$  field of view at a sampling of 0$^{\prime\prime}$.116  pixel$^{-1}$. The observations were made during the telescope commissioning period and good image quality with 0$^{\prime\prime}$.3--0$^{\prime\prime}$.4 seeing (FWHM of stellar images) was achieved. In this paper, we investigate the NIR properties of the host galaxy of 3C 324. Thanks to the high spatial resolution, we can directly compare the light distribution in the optical images obtained with the deep HST observations without seriously degrading the HST data. In section 2, we briefly describe the observations and the data reduction. We show the light distribution and color map of the host galaxy in section 3 and discuss its properties as a brightest cluster galaxy in section 4. We present our conclusions in section 5. 

\section{Observations and Data Reduction}

 The observations were made under stable weather condition on 1999 April 2 (UT), during the telescope commissioning period. In total, we observed the field for $\sim 1$ hour to investigate the luminosity function of the cluster galaxies (Kajisawa et al. 1999), with a typical seeing of $\ltsim$ 0$^{\prime\prime}$.5. For this paper, we extracted only the images taken in good seeing quality (FWHM $<$ 0$^{\prime\prime}$.37). The total net exposure time of the resultant image is 800 s and the r.m.s. noise level per pixel corresponds to a $K$-band surface brightness of $\mu_K$=21.4 mag arcsec$^{-2}$. 

 A description of the reduction procedure is given in Kajisawa et al. (1999). Since there were small variations of the bias level from frame to frame, we developed a procedure to construct a flatfield frame after removing the bias residual.  Otherwise, we followed the standard procedures in reducing NIR imaging data. The flux values were calibrated to those in the $K$ band by using the data of faint UKIRT standard star FS 27 ($K^\prime - K \sim 0.01$ mag) observed immediately after 3C 324 at a similar zenith distance. Using the model spectra of an old passively evolving galaxy at $z  =1.2$ and the transmission curve of the CISCO $K^\prime$ filter, we estimate the color of 3C~324 to be $K^\prime - K \sim 0.1$ mag. We then applied this correction to the photometric calibration, although there may be an uncertainty of $\sim 0.1$ mag, since the true color term of the telescope and instrument has not been well defined at this stage.

 We compare the NIR image with the deep WFPC2 images of the field of 3C 324. We retrieved the calibrated ${\rm F702W}$ and ${\rm F450W}$ images (PI: M. Dickinson; PID 5465 and 6553, respectively;see also Dickinson et al. 1995) from the Space Telescope Science Institute data archive. The total exposure times of these images are 64800 s (${\rm F702W}$) and 15000 s (${\rm F450W}$), and the ${\rm F702W}$ image is much deeper than that analyzed in Longair et al. (1995), Best et al. (1997), and Best et al. (1998b). We use $B_{{\rm F450W}}$ and $R_{{\rm F702W}}$ to denote the magnitudes measured in the STMAG system (HST Data Handbook Vol.1). STMAG is defined as $m_{ST} = -21.10-2.5 {\rm log} f_\lambda$ where $f_\lambda$ is expressed
in erg cm$^{-2}$ s$^{-1}$ \AA $^{-1}$.

\section{Light Distribution in the Host Galaxy}

\subsection{Images}

 Figure 1a shows the observed $K^\prime$ image of a 17$^{\prime\prime}$ $\times$ 17$^{\prime\prime}$ field centered on 3C 324. The host galaxy is clearly detected as a bright spheroidal galaxy, and surrounding objects which may be other cluster members are resolved. The total magnitude (MAG BEST value of the SExtractor output; Bertin, Arnouts 1996) of the host galaxy, obtained after separating the close companions, is $K=17.3$ mag.  Figure 1b shows a contour map of the $K^\prime$-band surface brightness distribution. For a comparison, we also show the original HST images taken in ${\rm F702W}$ and the ${\rm F450W}$ filters in figures 1c and 1d, respectively. While the optical images show elongated knotty structures, the light distribution in the $K^\prime$-band image of the galaxy is fairly smooth and round.

 Figure 2 shows the combined color image. The WFPC2 images were smoothed with a Gaussian kernel to match the FWHM of the stellar images to those of the Subaru $K^\prime$ image. The difference in the light distribution between the optical and NIR images is clearly seen. The peak of the NIR light clearly coincides with the gap in the optical structures, and thus also coincides with the position of the radio core (Best et al. 1998b), which implies that the powerful AGN indeed lies at the nucleus of the host galaxy. The gap in the optical structure may be due to a dust lane which hides the UV-optical emission of the AGN from direct view; it is very likely that we are seeing the obscuring structure from almost edge-on.

\subsection{Light Distribution and Near-Infrared Alignment Effect}

In order to investigate the intrinsic light profile of the host galaxy, we have to correct for contamination by the nearby companions. We removed them by replacing the pixels within 10-pixel (diameter) apertures centered on each object by an interpolation of the surrounding region. The surface brightness contours of the resultant image are shown in figure 3; any contamination by the close companions was removed well. The one-dimensional light profile obtained by isophotal-ellipse fitting to the resultant image is shown in figure 4. For a comparison, we also plot the light profile of a star in the same frame normalized at the peak.

 The galaxy has a smooth and regular light distribution, and the profile can be approximated by a de Vaucouleurs law with an effective semi-major axis (SMA) of 1$^{\prime\prime}$.3$\pm$ 0$^{\prime\prime}$.1 (11.2 kpc) and an effective surface brightness of $\mu_K$=20.3$\pm 0.2$ mag arcsec$^{-2}$, except for the central region affected by the effects of seeing and the outer region where sky noise dominates. We show the result of the fitting that was made for the SMA between 4 and 25 pixels (0$^{\prime\prime}$.46--2$^{\prime\prime}$.88).  The position angle of the fitted isophotal ellipse at large radius (at $\mu_K \gtsim 21$ mag arcsec$^{-2}$) is $\sim 90^\circ$, similar to the position angle of the optical structures ($\sim 100^\circ$, Longair et al. 1995) and also close to that of the radio axis ($\sim 71^\circ$); there is therefore a weak alignment effect also in the $K^\prime$-band image. The effective radius evaluated in our profile fitting is significantly smaller than the value of 2$^{\prime\prime}$.2 published in Best et al. (1998a). This may be due to the relatively large seeing size of the UKIRT image and contamination by the close companions shown in figure 1a.

 Since the fit with the de Vaucouleurs law shown in figure 4 is not perfect, and it may overestimate the surface brightness at large radius (consequently, overestimate the effective radius), we also performed a curve of growth analysis. The SMA of the ellipse that contains a half of the total flux is 0$^{\prime\prime}$.8, smaller than the value obtained in the light-profile fitting. The disagreement may be due to a contribution by the obscured AGN at the center of the galaxy and/or to the component that causes the alignment effect. We find that $\sim 25$\% of the total flux comes from the region within the central 0$^{\prime\prime}$.37 (= FWHM) SMA elliptical aperture. If the flux within this aperture is completely dominated by the AGN component alone, the half-light semi-major axis could be as large as $\sim 1$$^{\prime\prime}$.1.

 In order to evaluate the contribution from the `aligned component', we compared the observed $K^\prime$-band image with models of a smooth elliptical galaxy with an ideal de Vaucouleurs profile constructed  by using the IRAF ARTDATA package. We consider the two limiting cases: (i) the host galaxy indeed has an effective SMA of 0$^{\prime\prime}$.8 and the apparent large effective SMA is due to the contribution by the extended `aligned' component, and (ii) the surface brightness at SMA $\sim 1$$^{\prime\prime}$.3 is dominated by the host galaxy and the disagreement with the half-light SMA may be due to the contribution by the nuclear (point source) component and/or to an intrinsic light profile which deviates from a de Vaucouleurs law at large radius. For the first case, we construct a model galaxy with an effective SMA of 0$^{\prime\prime}$.8 and the same position angle and ellipticity as the observed best-fit ellipse at SMA = 0$^{\prime\prime}$.8. The model is convolved with a Moffat point spread function with FWHM of 0$^{\prime\prime}$.37 and normalized to the observed galaxy flux within a central 2-pixel-radius circle. For the second case, we use a model with SMA = 1$^{\prime\prime}$.3, normalized to the observations at the effective SMA. We then subtract the model images from the data to investigate the residual `aligned component' in the resultant images. These two limiting cases should bracket the range of the possible strengths of the aligned component.

 Figures 5 and 6 show the results. In panels (a) and (b) of both figures, we show the observed $K^\prime$-band image after removing the close companions (equivalent to figure 4) and the image of the adopted model galaxy to be subtracted from the observed one. In panels (c) and (d), we show the resultant images after subtraction and a version of these images smoothed with a 3-pixel boxcar. For a comparison, in panels (e) and (f) we show the HST images smoothed with a Gaussian kernel to the spatial resolution of the Subaru $K^\prime$-band image.  The crosses show the position of the light peak in the $K^\prime$-band image. In both cases, an aligned component is seen whose morphology shows excellent agreement with the optical images. As expected, it is more conspicuous for the case of the model with an effective SMA = 0$^{\prime\prime}$.8. The peak surface brightness of the residuals is $\sim 10$--$20$\% of the host galaxy peak in the case of SMA=0$^{\prime\prime}$.8 and a few \% for the case of SMA = 1$^{\prime\prime}$.3.

 It is difficult to tell which case is true from the observed data. The subtraction of the host galaxy seems to be good, but not perfect, in both cases. There is some diffuse positive residual in the northeastern part of the galaxy in figures 5c and 5d, which suggests that the galaxy may be more extended than the SMA = 0$^{\prime\prime}$.8 de Vaucouleurs profile, although it could be residuals from the subtracted close companion. At the same time, the counts in the northwestern part of the residual image are negative which may suggest the existence of a nuclear component. Although the residual appears to be fairly flat in figures 6c and 6d, the counts become negative at SMA $\gtsim 2$$^{\prime\prime}$, consistent with the fact that the half-light SMA is smaller than the effective radius obtained from the profile fitting.

 What is the origin of the aligned component seen in the $K^\prime$-band image? If it is the continuum emission associated with the rest-frame UV light seen in the HST images, we may constrain its nature by investigating the spectral energy distribution of the component. Alternatively, there is a possibility that it is dominated by the line emission of [S {\sc III}] $\lambda$9532 \AA . Rawlings et al. (1991) shows that the [S {\sc III}] emission line sometimes account for $\sim 10\%$ of the observed broad-band flux although there may be some variation in the [S {\sc III}] line strength among the powerful radio galaxies and the uncertainty in their line flux measurements seems to be somwhat large (15--50\%; their table 2).

\subsection{Two-Color Diagram}

 In order to constrain the origin of the NIR light, we also investigate the spatially resolved spectral energy distribution of the objects assuming that the $K^\prime$-band aligned component is dominated by the continuum emission. The degraded HST images, whose seeing was matched to that of the $K^\prime$-band image, are used for the purpose. We show the twelve positions where the photometry was performed with the used aperture (2-pixels radius) superposed on the $K^\prime$-band image in figure 7, and the resultant two-color diagram is shown in figure 8. \#1 corresponds to the peak of the $K^\prime$-band light distribution or the gap in the optical structure. \#2--\#5 sample the colors of the outer part of the host galaxy less contaminated by the UV-optical structure. \#6--\#9 correspond to the four knots in the optical images and \#10--\#12 are located at the positions of companion galaxies. The peaks of the knotty structures on the western side of the galaxy are slightly ($\sim 2$ pix) different between the ${\rm F450W}$ and ${\rm F702W}$ images and we put the aperture on the ${\rm F450W}$ peak. The sky level was determined well outside of the galaxy and was assumed to be constant over the galaxy in each band. The r.m.s. fluctuation of the sky level was evaluated with the same size of aperture over the 17$^{\prime\prime}$ by 17$^{\prime\prime}$ field avoiding the detected objects. We provide $2 \sigma$ upper limits for positions \#5, \#10, and \#12 in the $B_{{\rm F450W}}$ image.

 For a comparison, the colors of model galaxies with various ages observed at z = 1.2 are also plotted in figure 7. These models were calculated by using GISSEL 96 (Bruzual, Charlot 1993) with a Salpeter IMF and solar metallicity. We have also plotted the colors of a quasar at $z = 1.21$ mimicked by a power-law with $\alpha=-0.7$ ($f_\nu \propto \nu^\alpha$; Cristiani, Vio 1990). The effect of reddening evaluated using the Galactic extinction curve (Cardelli et al. 1989) as well as the Calzetti's formula (Calzetti 1997) for $E(B-V)=0.3$, is indicated by the arrows. 

 The host galaxy has a very red $R_{{\rm F702W}}-K$ color, as expected for the oldest passively evolving galaxies. It is thus likely that the bulk of the infrared light from the host galaxy comes from an old stellar population at an age of $\sim 2--4$ Gyr. It seems difficult to explain the observed red color by dust reddening alone, since a huge amount of reddening, $E(B-V) \sim 1$ mag, is needed, while at least one of the apertures, \#4, is located in the outer region of the galaxy away from the putative dust lane (see figures 2 and 8).  On the other hand, the observed $B_{{\rm F450W}}-R_{{\rm F702W}}$ colors at all positions in the host galaxy are considerably bluer than in the passive evolution models. These colors are better represented by the mildly evolving model with ongoing star formation. Ongoing star-formation activity is naturally expected from the existence of a significant amount of dust ($\sim 10^8$ $M_\odot$, Best et al. 1998c). The bulk of the stars is, however, formed at a very early stage, even in such a mild-evolution model, and only a very small amount of star formation activity remains at the observed epoch. Alternatively, the blue $B_{{\rm F450W}}-R_{{\rm F702W}}$ color may be due to a small amount of contamination by the `aligned' component, even at the outer region of the galaxies. The three companion galaxies (apertures \#10--\#12) have colors that are  rather consistent with the old passively evolving models.

 The $R_{{\rm F702W}}-K$ colors at positions \#2 and \#5, namely the southwestern side of the galaxy, are redder than those at \#3 and \#4 (the northeastern side). This may be due to reddening by asymmetrically-distributed dust. Alternatively, there is an excess of $K^\prime$-band light at $\sim 1$$^{\prime\prime}$ south of the nucleus (see figures 5d and 6d) which may be another red cluster member seen {\it behind} the host of 3C 324, and therefore further reddened by dust in the radio galaxy.

 At position \#1, there may be a contribution from reddened AGN light. Indeed, the colors of \#1 are also consistent with those of an AGN reddened by $E(B-V) \sim 0.6$, and it is difficult to distinguish between the mild evolution model and the reddened AGN in the two-color diagram. Spectroscopic observations are needed to resolve this issue; the redshifted 4000 \AA\ break may be observed at $\sim 9000$ \AA\ if the stellar light dominates. 

 The observed colors of the optical knots, \#6--\#9, are redder than the AGN and star-formation models. However, they agree well with the AGN spectrum reddened by $E(B-V) \sim 0.15$. Note that the effect of the underlying host galaxy on the $K^\prime$-band magnitude is negligible at positions \#8 and \#9 (figure 2). This agreement implies that scattered light from the hidden nuclear source provides a significant fraction of the observed flux. In fact, Cimatti et al. (1996) measured the polarization degree, $\sim 10 \%$ at 4000--8000 \AA . Since the structures at \#8 and \#9 are well separated from the host galaxy, they must be clouds of material formed as a consequence of the jet activity. Although Cimatti et al. (1996) argue that \#9 may be a blue star-forming companion galaxy, the identical colors of \#8 and \#9 suggest that they have the same origin, which is a consequence of the AGN activity.

 At \#6 and \#7, we must correct for the contribution from the host galaxy to obtain the colors of the `aligned' components, which we show by the open triangles denoted by \#6A and \#7A in figure 7. We performed photometry on the $K^\prime$-band image after removing the model host galaxy as described above. The $B_{{\rm F450W}}$- and $R_{{\rm F702W}}$-band magnitudes of the aligned component were obtained by subtracting the contribution from the host galaxy ($\sim 10--25 \%$) from the observed flux; we evaluate the host contribution assuming the $R_{{\rm F702W}}-K$ and $B_{{\rm F450W}}-K$ colors at \#1. The $R_{{\rm F702W}}-K$ colors at \#6A and \#7A are much bluer than those at \#8 and \#9, suggesting that the emission mechanism might be different between the knotty structure inside and outside the host galaxy.

 Assuming that the light of the optical knots is dominated by scattered radiation, the scattering mechanism can be determined from the observed spectra (e.g., Cimatti et al. 1996 and references therein). In the case of electron scattering, the observed spectrum has the same shape as the incident spectrum. Scattering by optically thin dust predicts a `blued' spectrum (Cimatti et al. 1993).
%
%
 We assume an incident spectrum represented by an $\alpha = -0.7$ power-law.  Models for scattering by optically thin dust predict that the reprocessed spectrum is bluer than the incident one, and to be a flat spectrum, $f_\nu \sim$ constant (Cimatti et al. 1993). The observed $B_{{\rm F450W}}-R_{{\rm F702W}}$ and $R_{{\rm F702W}}-K$ colors at \#8 and \#9 coincide with those of the incident spectrum (figure 7) with a small amount of reddening. On the other hand, a large amount of reddening is needed to match with the flat spectrum, and we must demand a rather peculiar distribution of dust in the knots at \#8 and \#9 (i.e., optically thin to nuclear light but with a large reddening along the line of sight). Electron scattering with a relatively small amount of reddening along the line of sight seems to be favored in explaining the colors at \#8 and \#9.

On the other hand, while the corrected $R_{{\rm F702W}}-K$ colors at \#6 and \#7 are as blue as the flat spectrum, the $B_{{\rm F450W}}-R_{{\rm F702W}}$ colors are $\sim 1$ mag redder than the model predictions. It seems to be difficult to explain the observed color at \#6 and \#7 by the simple electron- or thin-dust scattering models. We compare the colors at \#6A and \#7A with the predictions of nebular continuum emission (e.g., Aller 1987; Dickson et al. 1995). We consider two rather extreme cases: (i) N(He$^+$)/N(H$^+$) = 0.1 and N(He$^{++}$)/N(H$^+$) = 0 (`NC1' in figure 7), and (ii) N(He$^+$)/N(H$^+$) = 0. and N(He$^{++}$)/N(H$^+$) = 0.1 (`NC2'). An electron temperature, $T_{\rm e}$, of 10000 K is assumed. The colors become $\sim 0.1--0.3$ mag and $\sim 0.2--0.4$ mag bluer in $B_{{\rm F450W}}-R_{{\rm F702W}}$ and $R_{{\rm F702W}}-K$, respectively, if we consider the case with $T_{\rm e}$=20000 K. The observed colors can be explained by nebular emission with a small amount of reddening, or by a mixture of the nebular emission and the flat spectrum. If the observed light at \#6 and \#7 is dominated by nebular continuum emission, we can predict the fluxes of the hydrogen recombination lines.  Under Case B, the flux of the H$\beta$ line is predicted to be $\sim 1--2 \times 10^{-16}$ erg s$^{-1}$ cm$^{-2}$ within a 0$^{\prime\prime}$.4 aperture, depending on the ionization structure of helium.

 Cimatti et al. (1996) estimated that unpolarized radiation contributes 50--70 \% of the optical flux within a 3$^{\prime\prime}$.8 by 1$^{\prime\prime}$ aperture. Although they argued that this could be due to photospheric emission from OB stars, we note that it can be explain by nebular continuum emission at positions \#6 and \#7.  In fact, the amount of reddening required to explain the observed $B_{{\rm F450W}}-R_{{\rm F702W}}$ color of the optical knots predicts far too red a $R_{{\rm F702W}}-K$ color ($R_{{\rm F702W}}-K$ $\sim 5.5$--6); the observed optical continuum emission of the `aligned' component cannot be dominated by a reddened starburst at \#6 and \#7.

\section{Properties as a Brightest Cluster Galaxy}

In this section, we discuss the near-infrared photometric properties of the 3C 324 host galaxy as the brightest cluster galaxy (BCG).

Arag\'on-Salamanca et al. (1998) showed that BCGs at $z = 0$--0.9 have luminosities consistent with the no-evolution prediction, while other giant ellipticals have some brightening as expected for passive evolution. Since the no-evolution model is unphysical because the galaxies must dim with cosmic time as the stars age, they argue that some merging has occurred in the BCGs since that epoch. Since the expected luminosity difference between no-evolution and passive evolution models increased with the redshift, the 3C 324 host galaxy could put stronger constraints on the issue.

Figures 9a and 9b show the apparent rest-frame $K$-band magnitudes of the BCGs in a 50-kpc aperture against redshift following Arag\'on-Salamanca et al. (1998). We plot the 3C 324 host galaxy by the star; the applied $k$-correction is $-0.69$ mag, which has been evaluated from Kodama and Arimoto's (1997) model spectrum of a giant elliptical galaxy. Clearly, it lies at the extension of the sequence of BCGs at $z = 0$--0.9 and this behavior is not consistent with the prediction of the passive evolution models. If $q_0=0.5$ is assumed, the 3C 324 host galaxy is even less luminous than the no-evolution prediction. The intrinsic luminosity is fainter than this since there is a contamination by the aligned component although it may be smaller than 0.1 mag (10\%).  The 3C 324 host may also experience merging or assembly events that have doubled its stellar mass since the observed epoch. 

 We also plot the other three BCGs and candidates at $z> 1$, namely the two spectroscopically confirmed brightest red galaxies in the regions of the CIG J0848+4453 at $z=1.27$ (ID 65 with $K=18.11$; Stanford et al. 1997), RX J0848.9+4452 at $z=1.26$ (ID 1 with $K=16.72$; Rosati et al. 1999), and the object ``G1'' with $K=17.23$ in the cluster near B2 1335+28 at $z \sim 1.1$ (Tanaka et al. 1999). Note that the magnitude values quoted from Stanford et al. (1997) and Rosati et al. (1999) are those with 2$^{\prime\prime}$.4 and 2$^{\prime\prime}$ apertures, respectively, and they
should be brighter in the 50-kpc aperture; a correction of $\sim -0.6$ mag (assuming the same light profile as the 3C 324 host galaxy) may be needed for a direct comparison with other data. The applied $k$-correction is $-0.68$ mag at $z=1.26$ and 1.27 and $-0.70$ mag at $z=1.1$. The apparent magnitudes of the BCGs from Stanford et al. and Tanaka et al. are fainter than the predictions of the passive evolution models for both the $q_0=0$ and $q_0=0.5$ cases.



\section{Conclusions}

 The host galaxy of the powerful radio galaxy 3C 324 was observed with the Subaru telescope under good seeing conditions. The host galaxy is clearly resolved and seen to be a spheroidal galaxy well approximated by a de Vaucouleurs profile. The effective (half-light) radius evaluated from profile fitting is 1$^{\prime\prime}$.3 (11.2 kpc), which is about half the value previously published in the literature, while a curve of growth analysis produces a value of 0$^{\prime\prime}$.8. After subtraction of the model galaxies, we clearly detect the `aligned component' in the $K^\prime$-band image. The disagreement between the effective radius obtained in the profile fitting and the half-light radius in the growth-curve analysis may be due to this `aligned component' and/or to a contribution from the obscured AGN.

The peak of the $K^\prime$-band light coincides with the position of the radio core, which strongly implies that the engine of the powerful radio sources is indeed hosted at the nucleus of the giant elliptical galaxy. The NIR peak also corresponds to the gap in the rest-frame UV emission, which may be due to a dust lane. It is very likely that we see the obscuring structure from an almost edge-on view. 

 The host galaxy has a very red $R_{{\rm F702W}}-K$ color and the near-infrared light of the galaxy is likely to be dominated by an old stellar population, while the relatively blue $B_{{\rm F450W}}-R_{{\rm F702W}}$ color suggests that there may be some small amount of star-formation activity.

The colors of the 'aligned' components located inside the host galaxy, which are obtained after subtracting the host component, may be explained by nebular continuum emission with a small amount of a dust while those outside the host galaxy are better modeled by optically-thin dust scattering of the nuclear light.

\vspace{0.5cm}

The authors are indebted to all members of the Subaru Observatory, NAOJ, Japan. We thank Nobuo Arimoto for kindly providing the Kodama and Arimoto evolutionary synthesis models. TY thanks Takashi Murayama for useful discussions.  We thank Dr. Marc Dickinson, the referee, for the invaluable comments. This research was supported by grants-in-aid for scientific research of the Ministry of Education, Science, Sports and Culture (08740181, 09740168). This work was also supported by the Foundation for the Promotion of Astronomy of Japan.  This work is based in part on observations with the NASA/ESA Hubble Space Telescope, obtained from the data archive at the Space Telescope Science Institute, U.S.A., which is operated by AURA, Inc.\ under NASA contract NAS5--26555. The Image Reduction and Analysis Facility (IRAF) used in this paper is distributed by National Optical Astronomy Observatories. U.S.A., operated by the Association of Universities for Research in Astronomy, Inc., under contact to the U.S.A. National Science Foundation.
\clearpage

\section*{References}
\small

%

\re
Aller, L. 1987,  Physics of Thermal Gaseous Nebulae, Astrophys. Space Sci. Library Vol.112, (Leidel, Dordrecht)

\re
Aragon-Salamanca A. , Baugh C. M. \& Kauffmann G.  1998, MNRAS 
297, 427

\re
Bertin E., Arnouts S. 1996, A\&AS 117, 393



\re 
Best P. N. Longair M. S., R\"oettgering J. H. A. 1997, MNRAS
292, 758

\re
Best P. N., Longair M. S., R\"oettgering, H. J. A. 1998a, MNRAS
295, 549

\re
Best P. N., Carilli C.  L., Garrington S. T., Longair M. S.,
Rottgering H. J. A. 1998b, MNRAS 299, 357

\re
Best P. N., R\"oettgering H. J. A., Bremer M.N., Amatti A., Mack K.-H., Miley G.K., Pentericci L., Tilanus R.P.J., van der Warf P.P., 1998c, MNRAS 301, L15 

\re
Bruzual A. G., Charlot S.  1993, ApJ 405, 538

\re
Calzetti D. 1997, The Ultraviolet Universe at Low and High Redshift : Probing the Progress of Galaxy Evolution,  ed. W. H. Waller et al. (Amrican Institute of Physics: New York), 403

\re
Cardelli J. A., Clayton G. C., Mathis J. S. 1989, ApJ 345, 245 

\re
Chambers K. C., Miley G. K., van Breugel W. 1987, Nature 329, 604

\re
Cimatti A., di Serego-Alighieri S., Fosbury R. A. E., Salvati 
M., Taylor D. 1993, MNRAS 264, 421 

\re
Cimatti A. , Dey A. , Van Breugel W., Antonucci R., Spinrad 
H.  1996, ApJ 465, 145

\re
Cristiani S., Vio R. 1990, A\&A 227, 385

\re
Dickinson M., Day A.,  Spinrad H., 1995, Galaxies in the Young Universe, eds. H. Hippelein, K. Meisenheimer, \& H.-J. Roser (Springer Verlag: Berlin), 164. 

\re
Dickinson M. 1997, The Early Universe with the VLT.  ed. 
J. Bergeron (Springer Verlag: Berlin), 274

\re
Dickson R., Tadhunter C., Shaw M., Clark N., Morganti R. 1995,
MNRAS 273, L29

\re
Dunlop J. S., Peacock J. A. 1993, MNRAS 263, 936

\re
Fernini I. , Burns J. O., Bridle A. H., Perley R. A. 1993, AJ
105, 1690

\re
HST Data Handbook, Ver3.1 1998, http://www.stsci.edu/documents/data-handbook.html

\re
Kajisawa M,  et al. 1999, PASJ in press.

\re
Kodama T., Arimoto N. 1997, A\&A 320, 41

\re
Longair M.  S., Best P. N., R\"ottgering H. J. A. 1995, MNRAS
275, L47

\re
McCarthy P. J., van Breugel W. , Spinrad H., Djorgovski, S. 1987, ApJL 321, L29 

\re
Motohara K., et al. 1998, Proc. SPIE 3354, 659

\re
Rigler M. A., Lilly S. J., Stockton A., Hammer F., Le Fevre O. 1992, ApJ 385, 61 

\re
Rosati P., Stanford S. A., Eisenhardt P. R., Elston R., Spinrad 
H., Stern D., Dey A. 1999, AJ 118, 76


\re
Stanford S. A., Elston R., Eisenhardt P. R., Spinrad H., Stern D., Dey A. 1997, AJ 114, 2232

\re
Tanaka I. , Yamada T. , Arag\'on-Salamanca A. , Kodama T. , Ohta 
K., Miyaji T., Arimoto N.  1999, ApJ in press.


\begin{fv}{1}{18pc}%
{$K^\prime$ images of 3C 324 taken with the Subaru telescope (panel a) and the isophotal contour map for the image after boxcar smoothing with 3$\times$3 pixels (panel b). The lowest contour in panel (b) corresponds to the 1$\sigma$ noise level of the sky {\it before} the smoothing; the contour interval is 0.5 mag arcsec$^{-2}$. The images HST/WFPC2 taken in the ${\rm F702W}$ (panel c) and the ${\rm F450W}$ filters (panel d) are also shown. The box spans 17$^{\prime\prime}$ in panels (a), (c), and (d) and 10$^{\prime\prime}$ in the panel (b).  }
\end{fv}

\begin{fv}{2}{18pc}%
{Combined $B_{{\rm F450W}}R_{{\rm F702W}}K^\prime$ three-color image of 3C 324. The optical images taken with HST have been Gaussian-smoothed to match the PSF to that of the $K^\prime$-band image.}
\end{fv}

\begin{fv}{3}{18pc}%
{Contour map of the $K^\prime$ image of 3C 324 after removing the close companions. The levels of the contours are the same as in figure 1b. The box spans 5$^{\prime\prime}$.  }
\end{fv}

\begin{fv}{4}{18pc}%
{One-dimensional light profile of the 3C 324 host after removing the close companions. The light profile of a star in the frame is also shown with crosses. The horizontal bar indicates a radius of twice the half width of the half maximum. The dashed line shows the best-fit de Vaucouleurs profile, fitted between 4-25 pix (0$^{\prime\prime}$.46--2$^{\prime\prime}$.88).  }
\end{fv}

\begin{fv}{5}{18pc}%
{Observed $K^\prime$-band image after removing of the close companions (panel (a)) and the image of the adopted model galaxy with an effective semi-major axis of 0$^{\prime\prime}$.8 is shown in (b). The resultant image after subtraction of the model host in shown in (c) and a boxcar-smoothed version of this is shown in (d). A NIR alignment effect is clearly detected. For comparison, we show the Gaussian-smoothed HST images in panels (e) and (f). The crosses show the position of the light peak in the $K^\prime$-band image.  }
\end{fv}

\begin{fv}{6}{18pc}%
{ Same as figure 5 but for the model with an effective semi-major axis of 1$^{\prime\prime}$.3. }
\end{fv}

\begin{fv}{7}{18pc}%
{Positions and apertures (2-pixel radius) for the photometry presented in figure 8.}
\end{fv}

\begin{fv}{8}{18pc}%
{Two-color diagram of the various positions (\#1--\#12) shown in figure 7. The colors of the `aligned' component after subtracting the host galaxy models with $r_{\rm e}=0$$^{\prime\prime}$.8 (upper ones) and $r_{\rm e}=1$$^{\prime\prime}$.3 (lower ones) are also plotted by the open triangles, denoted as 6A and 7A. The tracks of the passive evolution model with a 1-Gyr initial starburst (long dashed line) and the mild evolution model with an exponential star-formation decaying time scale of 0.5 Gyr (short dashed line) with various ages are shown for comparison. The solid tick marks on the tracks correspond to ages of 2, 2.5, 3, 3.5, and 4 Gyr; the reddest one is the oldest. The colors of constant star-formation models at ages of 0.01 Gyr and 0,1 Gyr are shown by the crosses. The colors of power-law spectra with $\alpha=-0.7$ and $\alpha = 0$ are also indicated by the star labeled  `AGN' and the asterisk labeled as `$f_\nu$ = const.', respectively. NC1 and NC2 refer the colors of the nebular thermal continuum emission (see text for the detail). The arrow shows the effect of reddening calculated using the galactic extinction curve by Cardelli et al.(1989) as well as the Calzetti's (1997) formula for starburst galaxies with $E(B-V)=0.3$. }
\end{fv}

\begin{fv}{9}{18pc}%
{$k$-corrected apparent magnitude of the 3C 324 host galaxy (star) shown together with those of the brightest cluster galaxies studied in Arag\'on-Salamanca et al. (1998) (filled circles) and those in the clusters at $z > 1$ (open circles; see text). Prediction of the case of no evolution as well as the passive evolution models are shown for $H_0$ = 50 km s$^{-1}$ Mpc$^{-1}$ and $q_0$ = 0.5 (panel a) and 0.0 (panel b).}
\end{fv}

\clearpage

\begin{figure} 
\psfig{file=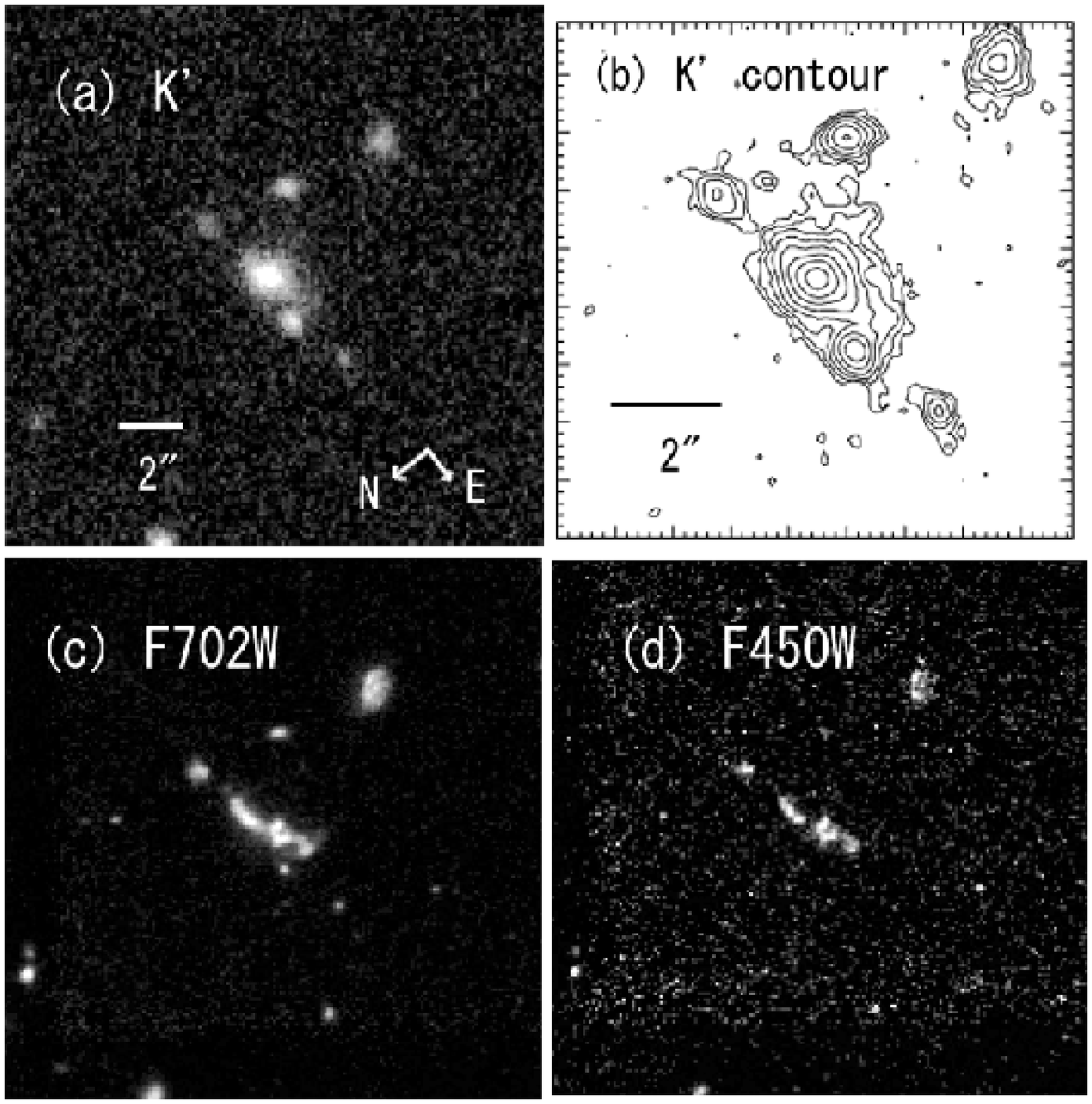,width=1.0\textwidth}
\caption[fig1.ps]{}  
\end{figure}

\clearpage

\begin{figure} 
\psfig{file=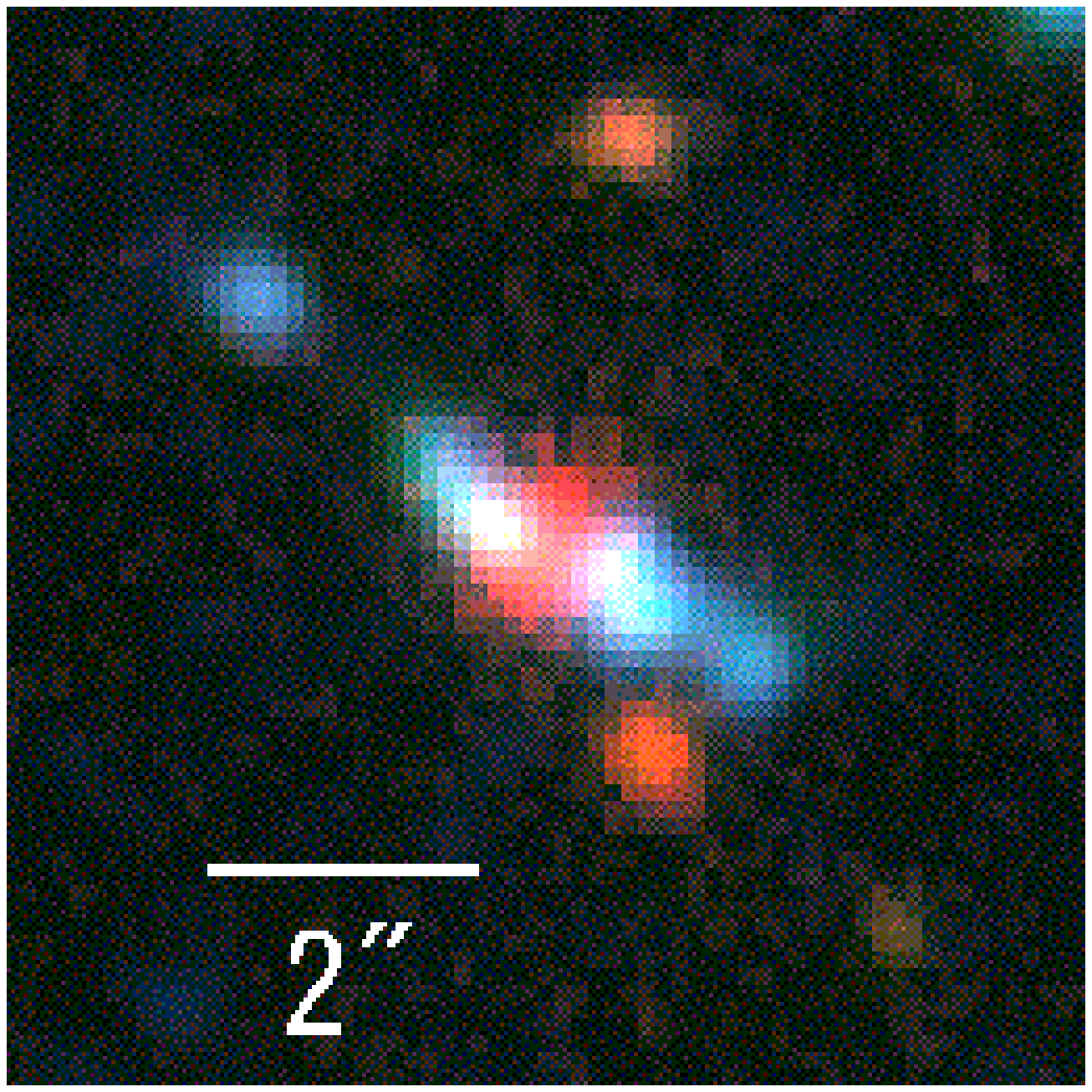,width=1.0\textwidth}
\caption[fig2.ps]{}  
\end{figure}

\clearpage

\begin{figure} 
\psfig{file=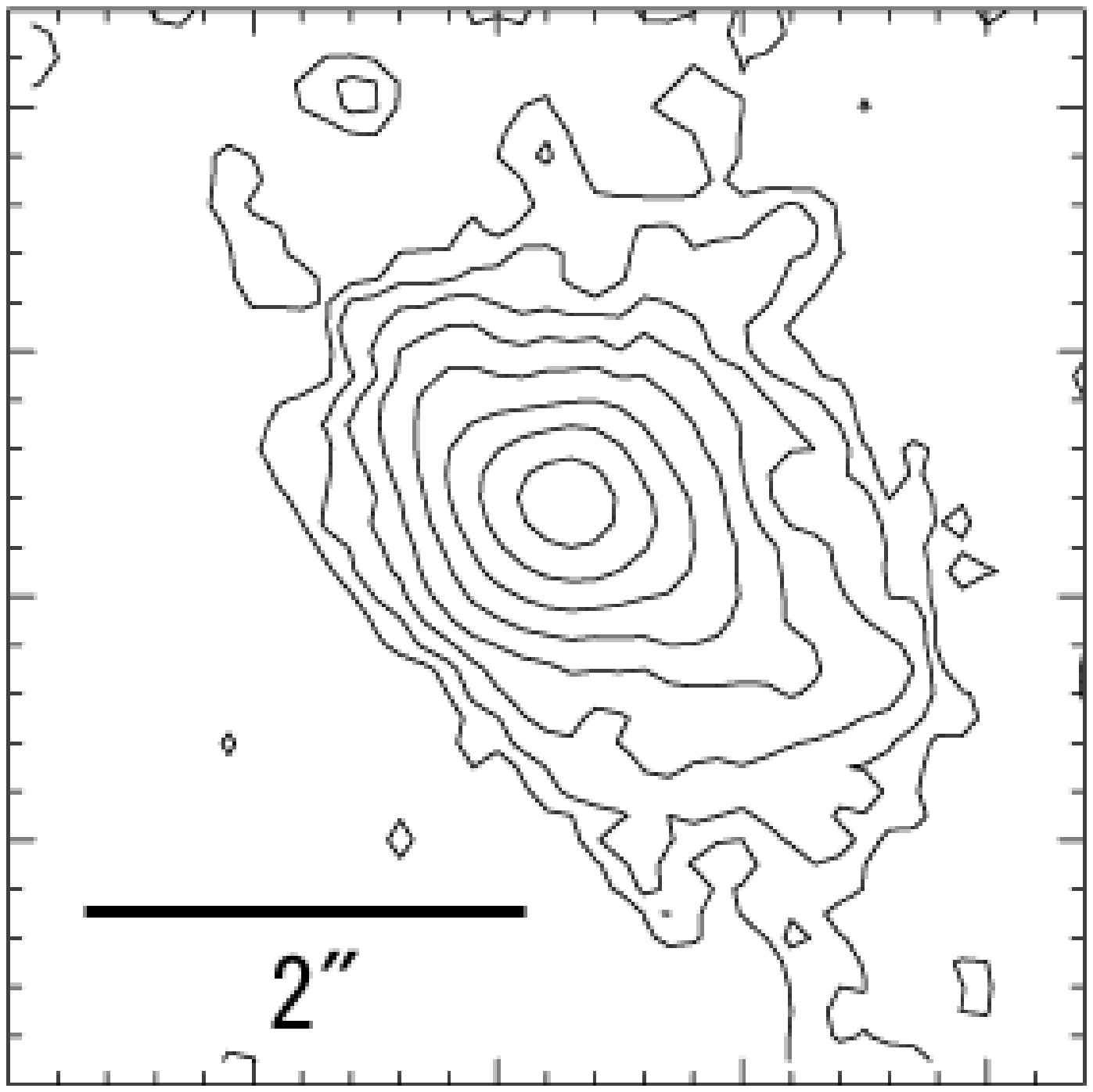,width=1.0\textwidth}
\caption[fig3.ps]{}  
\end{figure}

\clearpage

\begin{figure} 
\psfig{file=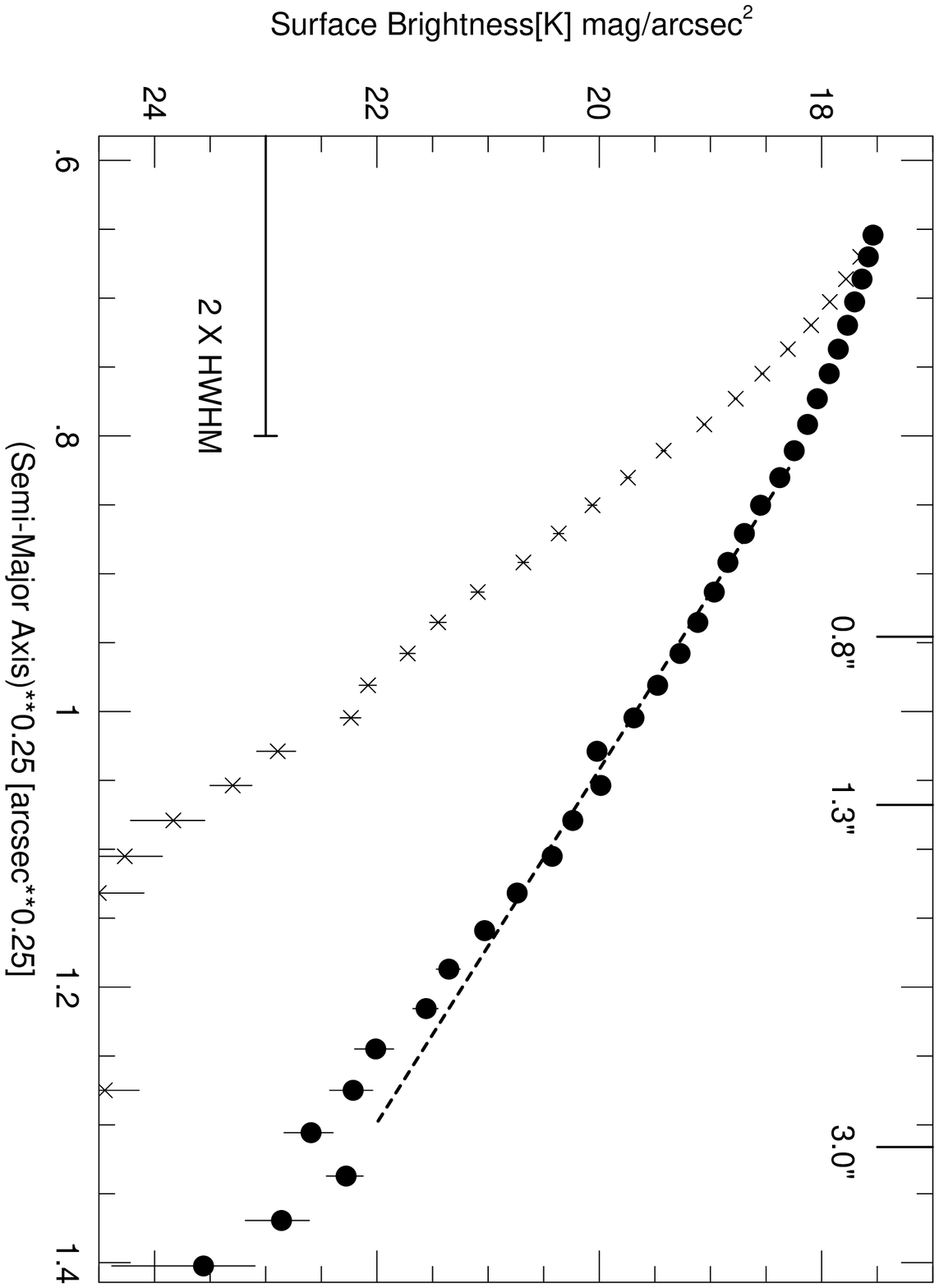,width=1.0\textwidth}
\caption[fig4.ps]{}  
\end{figure}

\clearpage

\begin{figure} 
\psfig{file=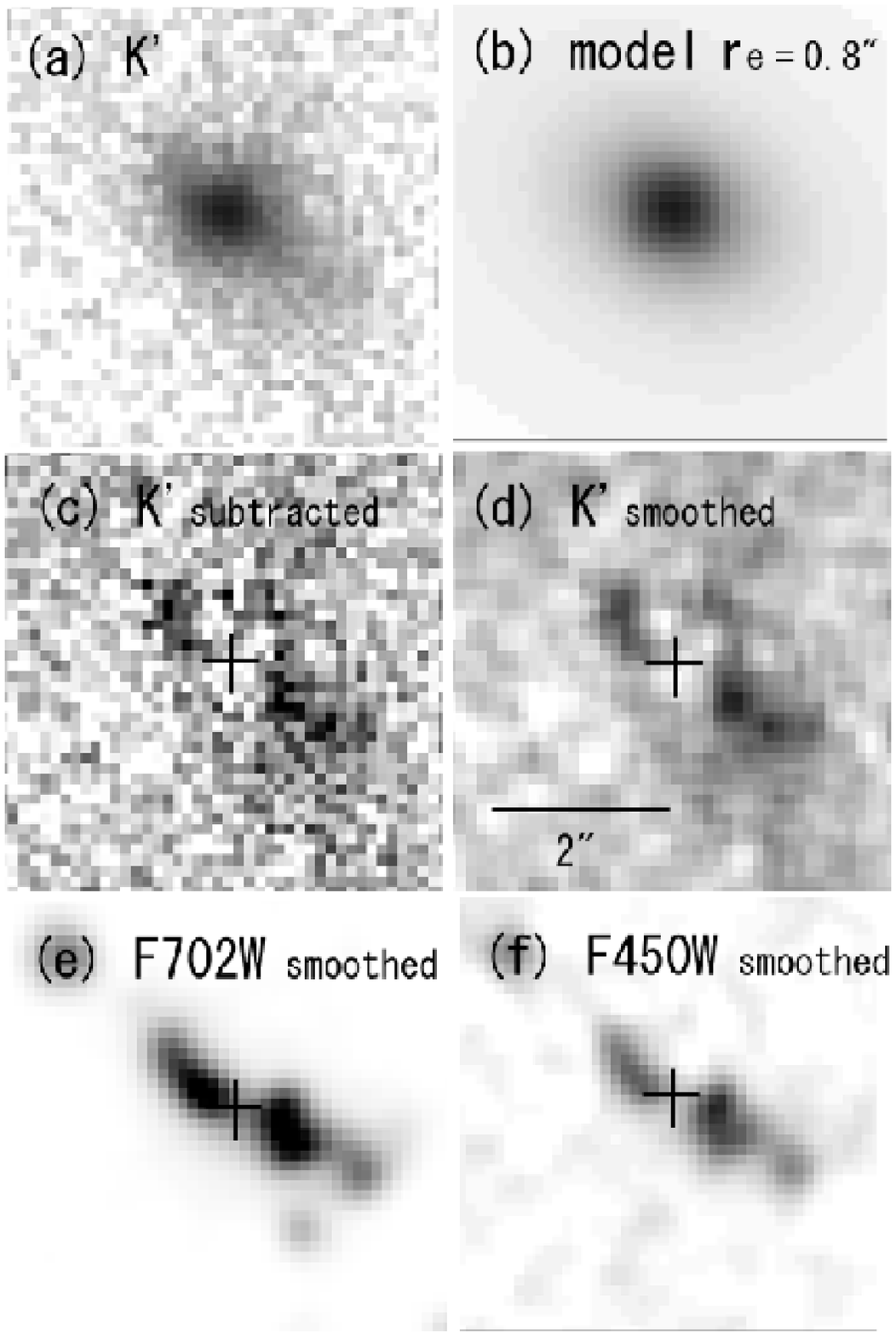,width=1.0\textwidth}
\caption[fig5.ps]{}  
\end{figure}

\clearpage

\begin{figure} 
\psfig{file=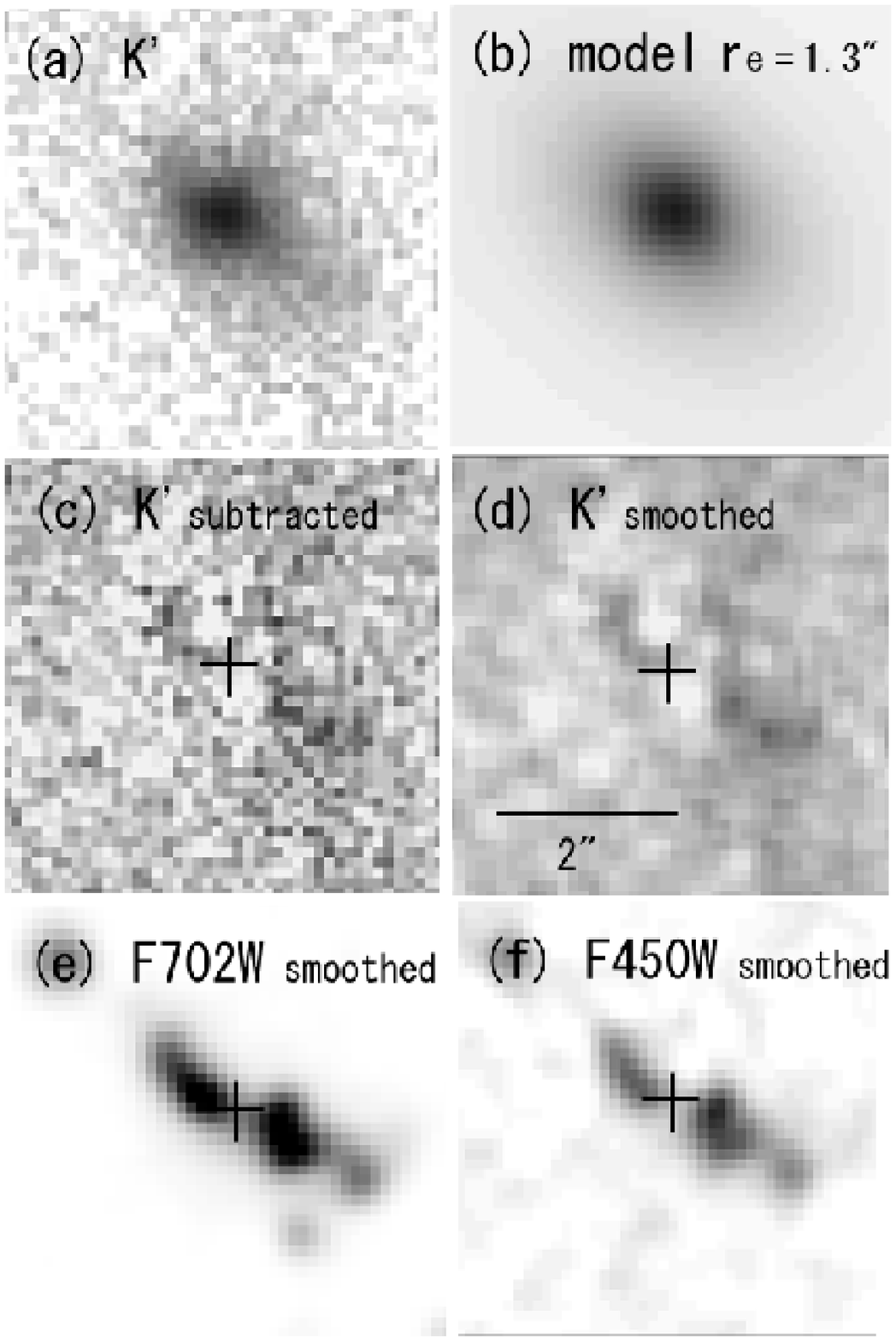,width=1.0\textwidth}
\caption[fig6.ps]{}  
\end{figure}

\clearpage

\begin{figure} 
\psfig{file=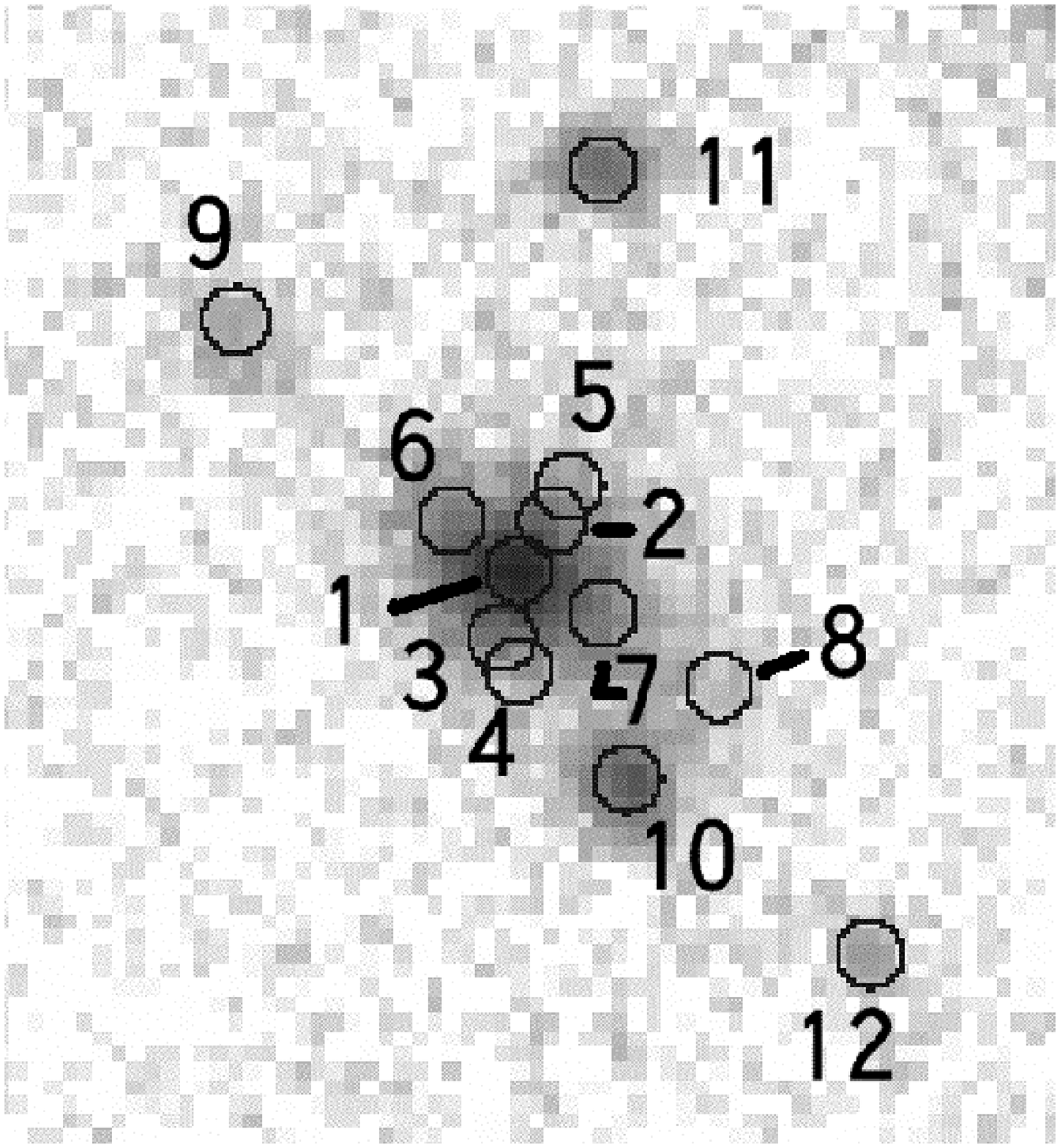,width=1.0\textwidth}
\caption[fig7.ps]{}  
\end{figure}

\clearpage

\begin{figure} 
\psfig{file=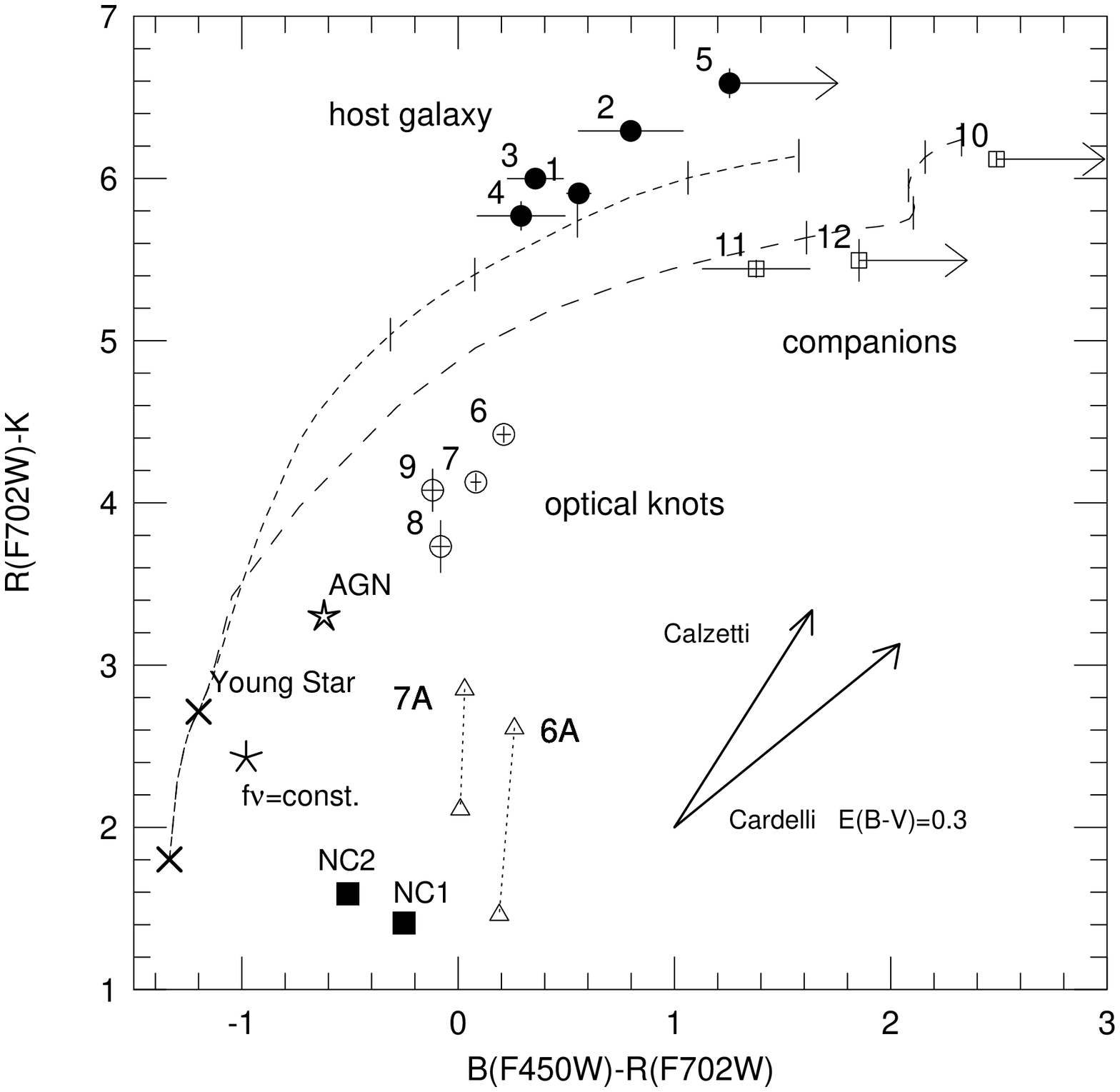,width=1.0\textwidth}
\caption[fig8.ps]{}  
\end{figure}

\clearpage

\begin{figure} 
\psfig{file=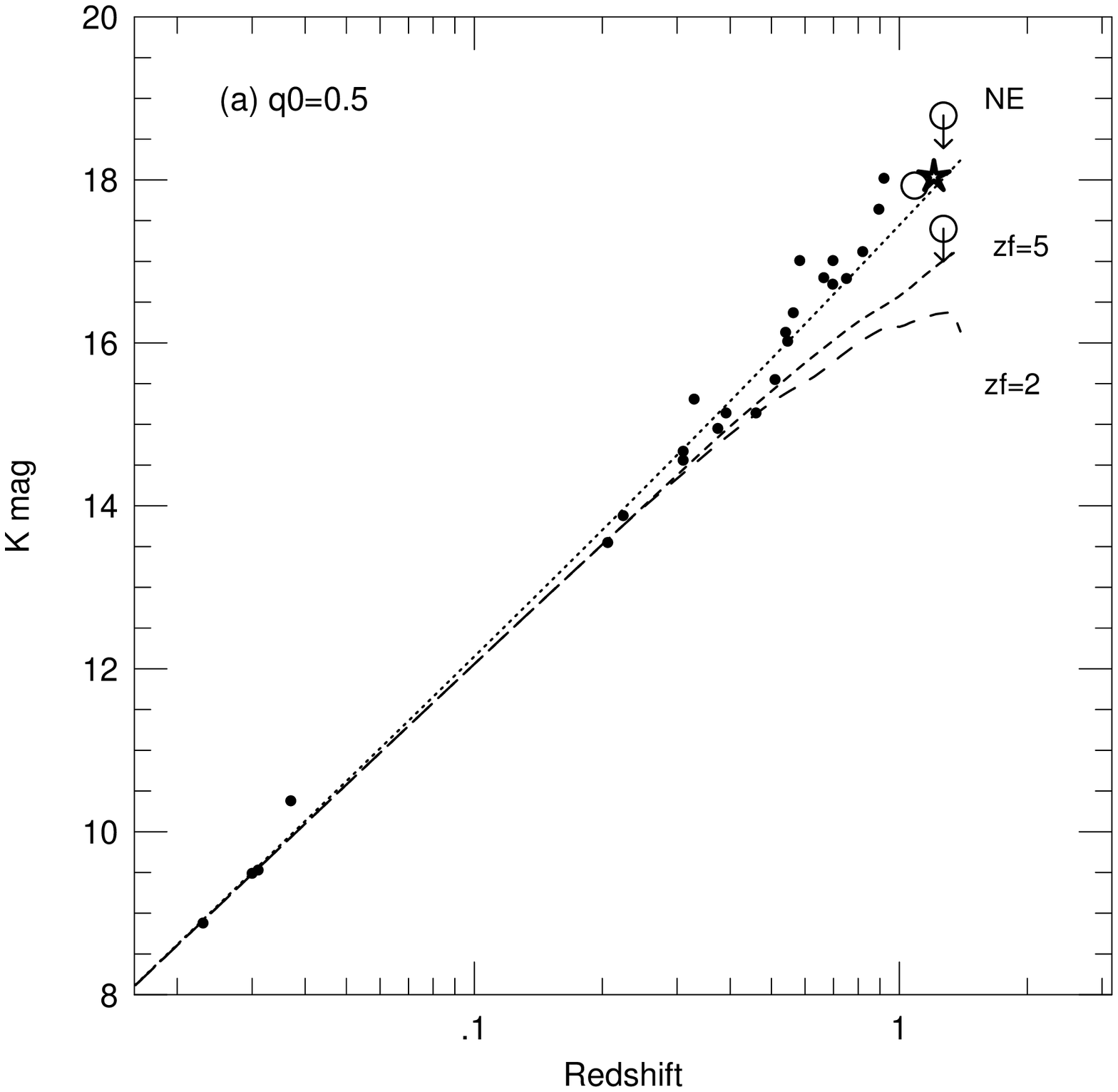,width=1.0\textwidth}
\caption[fig9a.ps]{}  
\end{figure}

\clearpage

\begin{figure} 
\psfig{file=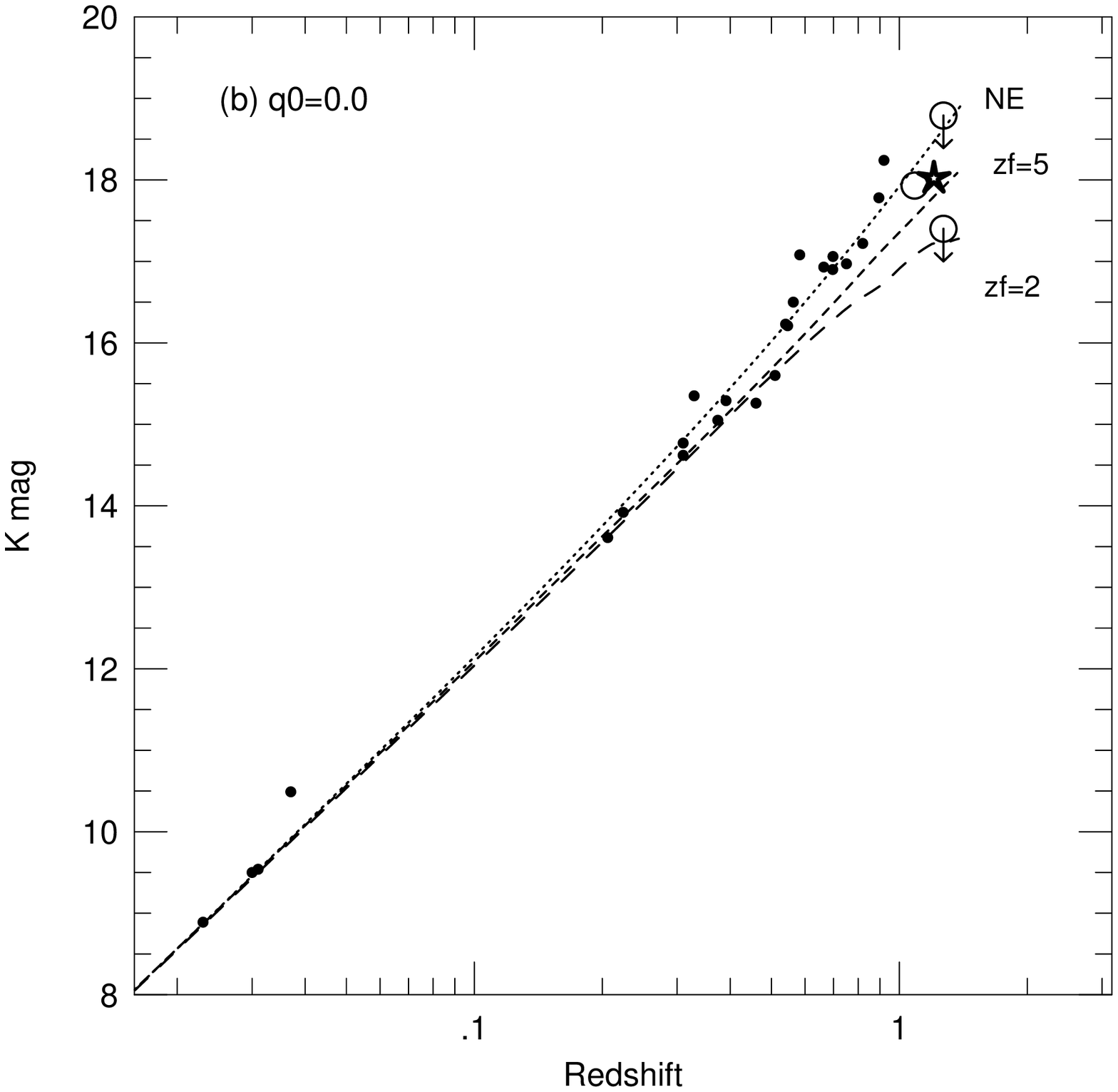,width=1.0\textwidth}
\caption[fig9b.ps]{}  
\end{figure}

\end{document}